# Technical paper: Collecting classroom video on a budget

Renee Michelle Goertzen
Department of Physics, University of Maryland, College Park, MD 20742, USA

Abstract. This paper describes low-cost techniques used to collect video data in two different tutorial classrooms – one in which the recording equipment is permanently installed and one in which it is temporary. The author explains what to do before, during, and after class in these two situations, providing general strategies and technical advice for researchers interested in videotaping tutorials or similar classrooms.

## I. INTRODUCTION

Videotaping students in the classroom supplies a wealth of data that observations or audio recordings cannot provide.[1,2] Classes using small-group collaborative learning activities such as tutorials are well suited for videotaping. This paper describes low-cost techniques that were used to collect video data in two tutorial classrooms. One of these classrooms had permanently installed equipment and one classroom used video equipment temporarily. I explain what to do before, during, and after class in these two situations, providing general strategies and technical advice for researchers interested in videotaping tutorials or similar classrooms.

I collected video as part of a project whose goal was to better understand the teaching done by physics graduate students using tutorials in introductory physics courses. Teaching assistants (TAs) at two institutions, the University of Colorado, Boulder (CU-B) and the University of Maryland, College Park (UM), participated in the study. In each case, the TAs were teaching students in introductory courses that had a one-hour TA-led discussion section each week. During the discussion sections, the students worked through tutorials, which are worksheet-based group learning activities,[3,4] instead of participating in typical problem-solving recitations.

At both institutions, TAs teach tutorials in rooms set aside for this purpose. The rooms have seven or eight tables that each seat four or five students. Two instructors teach each tutorial by circulating around the room assisting students. Every student has his or her own worksheet on which to write, and there is a large sheet of paper in the center of each table on which to write or draw for the whole group.

## II. BEFORE CLASS: SELECTING AND PREPARING THE EQUIPMENT

### A. Choosing equipment

I chose the videotaping system described here for its low cost and portability. All of the equipment described here is marketed to the average consumer rather than video professionals. This kind of system is appropriate to the researcher considering using video for the first time. A minimum requirement would be one camera and microphone, which together would cost around $300. (For a detailed discussion on setting up a permanent system with higher video quality, see Kung, Kung, & Linder.[5])

Video technology changes rapidly, so it is not expected that all the equipment recommended here will be available for other researchers. While the particular equipment selections are listed as examples, the reasons for each choice are discussed in detail so that the reader can use this analysis to guide selections among the equipment options that are currently available.

### *1. Video cameras*

I used four different video camcorders: a Canon ZR500, a Canon ZR800, a Canon Elura 100, and a Samsung SC-D365, all of which recorded to miniDV tapes. I chose this format over one that records directly to a hard drive because I required a system that did not need lengthy downloading between videotaping sessions and because the tape can act as an automatic hard copy backup. In addition, the popularity of the miniDV format ensures a wide range of camera options and increases the likelihood that the format will be long-lived. I rejected camcorders recording to mini DVD, flash memory, or hard drive because those formats are more difficult to edit and produce lower quality video.[6]

The cameras' hardware features affected my choice. I sought cameras with a microphone jack, a headphone jack, and FireWire transfer capability. Microphone and headphone jacks are not universal on consumer cameras, but acceptable audio recording of individual groups in a busy room requires a microphone, and a headphone jack is extremely useful for checking sound quality during taping and for viewing videos from the camera. FireWire is the standard connection for transferring video from cameras to computers. It is much faster than a USB connection and I have used it with a variety of miniDV cameras and computers. Selecting for these features narrows the camera choices considerably and I was unable to purchase cameras with headphone jacks.

I would have preferred cameras that only loaded from the side because bottom-loading cameras must be detached from the tripod before the tape can be changed. Unfortunately, side-loading cameras are now uncommon. I solved this problem by building a bracket to mount the camera off-center on the tripod so that I could change the tape without removing it from the tripod.

### *2. Microphones*

Microphones need to pick up sound from the desired directions and distinguish the immediate group from the noisy environment. A microphone's polar pattern maps the directions from which it receives sound. (See Rumsey & McCormick for further discussion.[7]) A microphone with an omnidirectional polar pattern was easiest to use because it picked up sound from all directions. I have also successfully used cardioid microphones, but they required vertical placement in the center of a group's table because they only picked up sound from the region above each microphone.

*(a) Microphones for portable use*

I have used two varieties of portable microphones: a Sony Omnidirectional ECM-F01 Microphone and Sony WCS-999 cordless lapel microphones. They were both easily attached and removed and could be placed in the center of the table to record all participants in the group. I checked for dead batteries frequently to avoid the loss of audio data. The wireless microphones could broadcast to their receivers on several channels; I found that certain channels picked up more outside noise than others did.

*(b) Microphones for permanent installation*

The miniDV cameras I used required powered microphones, but I wanted to avoid battery-powered microphones at UM because we permanently installed them and regularly checking battery strength would become inconvenient. These cameras would accept unpowered signals from dynamic microphones, so I chose two Behringer XM8500 Cardioid microphones. These microphones did not need batteries, but they had a cardioid shaped polar pattern, which meant that they needed to be oriented vertically when recording.

The Behringer microphone had an XLR connector, which is common for professional quality microphones. Because the miniDV cameras had miniplug connectors, as most camcorders do, an XLR to miniplug adaptor was needed to connect the two.

### 3. Audio recorders

In some cases, I made simultaneous audio and video recordings to minimize the chance that data would be lost. For this purpose, I used a battery-powered digital recorder, the Olympus DS-2 digital audio recorder. It was convenient because of its small size and its ability to record up to 22 hours of audio data.

## B. Setting up equipment in the classroom

### 1. Choosing which table to observe

Each classroom had six or seven student groups. Taping six groups at once would have required great coordination, and the richness of video data made huge numbers of tapes unnecessary. I selected two tables to tape in each classroom, but the reasons for the choice at each institution were different.

Taping tutorials at UM is an established practice and so we have permanently installed microphones and cameras at two tables in the tutorial room. (The placement of the equipment is discussed in the next two sections.) We find that students rarely change the group with which they work or the table at which they sit (although neither the groups nor the seating is assigned) and thus the same group is often taped during the entire semester. This provides us with an opportunity to become more familiar with a particular group and to build a deep understanding of their practice.

TAs at CU-B taught tutorials in multiple rooms and so the equipment needed to be portable and not permanently installed. This had the advantage of allowing us to tape a wider variety of groups. The disadvantage of moving cameras frequently was that microphones, wires, and power cords were more conspicuous.

### 2. Choosing the optimal camera location

A primary goal I had when taping was to minimize the obtrusiveness of the video recording, because I think that what people perceive they are doing affects what they actually do. That is, I wanted the students and the TAs to focus mainly on what they were teaching and learning, and minimally on what the camera might have been recording. As a result, I placed the cameras and tripods between twelve and fifteen feet away from the table being taped. It was then possible to zoom in the camera so that just one table was on the screen. I positioned the cameras higher than eye level to minimize the possibility that individuals would sit or walk in front of the camera. The camera was raised by using a tripod that extended to about six feet or by placing the tripod on a table. If I was using a wired microphone, I tried to place the cord around the perimeter of the room or run it up to the ceiling and across to the camera.

### *3. Choosing the optimal microphone location*

One of the greatest challenges I faced was determining which microphone would work in a noisy classroom environment. I chose the microphones for use at CU-B because I could easily move them. The Omnidirectional microphone, which is slightly larger than a credit card, was easily taped to the center of the table. I secured its cord to the edge of the tabletop and then across the floor to the camera. The lapel microphones were especially convenient because I could tape the small, quarter-sized microphone to the center of the table and position the cord so that the battery-powered transmitter was under the table. The wireless receiver for these microphones was plugged into the camera.

At UM, we use handheld vocal microphones, which we have inserted into holes drilled in the middle of the tables. The top sphere sits flush with the tabletop and the long handle extends under the table. This positions the microphone vertically so that it picks up sounds in all directions above the tabletop, which is an important concern with a microphone that has a cardioid shaped range. We have covered the microphone heads with plastic cages to protect them from damage.

### *4. Using an audio recorder*

I decided to make simultaneous audio recordings when opportunities to tape were limited. At CU-B, I was collecting data at a remote institution for only one month and could not retape any hours of unusable video in later tutorials or later semesters. I therefore used battery-operated digital recorders placed in the center of the tables. As a precaution, I began recording and then taped the recorders down so that the buttons were covered by tape. This was due to previous observations that students have a tendency to fiddle with equipment left on the table; I hoped this would prevent them from accidentally turning the recorders off.

### III. DURING CLASS: CONSENT AND TAPING

#### A. Obtaining permission from students

How a researcher introduces videotaping to the participants being taped is important, both to maintain ethical standards and to maximize the number of students or TAs who agree to participate. Once I obtained IRB approval, I requested permission from each tutorial TA. I then went to each tutorial classroom to announce the project to the students the first time I taped them.

### 1. *What participants need to know*

My primary goal when approaching participants to ask for informed consent was to convey respect for them and their right to learn in a comfortable environment. Therefore, one of the first things I told students is what would happen if someone did not give consent but still appeared on video. In that case, the videotaping would still occur, but if I wanted to use a tape of a nonconsenting student for research, I would blur the student so that he was unrecognizable and dub over any words he had spoken.

Participants needed to know a great deal of information before they could give their informed consent.[8] They needed to know what would happen if they did not participate: in the case described here, I explained that students' grades would not be affected and that the instructors (both TAs and lecturers) would not know whether the students had chosen to participate until after the end of the semester. If they did not chose to participate and did not wish to appear on video, I advised students that they could sit at a table that was not being taped. I never required them to do so, because I did not want to foster the perception that I was more interested in students who were willing to participate. The participants also needed to know how the video could be used: whether names (first or last) would be disclosed and where the video might be shown. (Options can include showing it only to researchers involved with the project, to researchers at conferences, or in publications.)

### 2. *Setting the tone*

I found that the fraction of people who were willing to be videotaped depended greatly on the tone of the researcher's explanation. If the explanation were confusing or uncertain, people might have felt uncomfortable agreeing to be videotaped. It was helpful to explain a discussion section project in the discussion section itself rather than in lecture, because students seemed more comfortable asking questions and because it was easier to convey an open and relaxed tone in a smaller setting.

It was important to know when researchers were allowed to look at the video collected. When I videotaped students in TA-led tutorials, the lecturers of the classes did not look at that video while they were still teaching that course because their opinions of their students (and thus their grades) could have been influenced by what they saw on the tape. On the other hand, if the lecturer had been the tutorial instructor, she could have watched the tape during the semester she was teaching because students would not have had an expectation that what happened in a classroom in her presence would be unknown to her.

## B. Taping

In line with the effort to minimize the effect of the taping on the students' and instructors' behavior, a researcher entered the room before the class started to turn on the cameras and then left during the recording. When multiple classrooms were used for taping, as happened during the data collection at CU-B, I sometimes needed multiple researchers to quickly remove and reposition the cameras at the end of one tutorial and the start of the next. I always made an effort to have the cameras positioned and turned on at least a few minutes before class started.

## IV. AFTER CLASS: PREPARING FOR ANALYSIS

### A. Digitization

Digitizing is the process of transferring the video on the tape to a digital form on a computer. It is necessary to digitize videos so that they can be edited, which includes changing video formats, making smaller video clips, adding captions, or obscuring nonparticipating students.

To digitize on a PC, I attached the video camera to the computer using a FireWire cable. I used the program Windows Movie Maker, which is bundled with Microsoft Windows. After starting to play the tape, I used the option "Capture video from video device" and could then capture the entire tape automatically or choose portions of the tape. Video digitized this way was in the .wmv format, which is hard to edit. I converted this to a .mov format using SUPER 2009 video converter, which is made by eRightSoft. This free program converts between many video formats. The interface on SUPER offered an overwhelming number of options, but if the defaults were used, only the output container (.mov) needed to be chosen. After that I dragged the .wmv file on to the empty encoding area and selected the option "Encode."

I have also digitized on a Mac using QuickTime Pro, which is the upgraded version of the free program QuickTime. (While QuickTime Pro is available for PCs, that version does not support digitizing videos.) After hooking up the camera to the computer with the FireWire and starting to play the tape, I chose the option "Make a new movie recording." This produced a video file in the .mov format.

### B. Editing

I chose my video clips before I transcribed them; this was a methodological choice,[1] but had the practical effect of allowing me to view larger quantities of video data before I began the time-consuming process of transcribing. Once I selected a section of the tape that I wished to examine further, I used QuickTime Pro to make it into a clip. The two pointers on the bottom slider bar were moved to mark the start and finish of the desired clip, and then I used the Edit menu command "Trim to Selection" and saved the clip as a smaller, separate file.

### C. Transcription

Transcribing an hour of tape took between four and ten hours, depending on the level of detail I included in the transcript (such as non-word sounds, pauses, or descriptions of gestures and tone changes). In my experience, voice recognition programs that automatically transcribe spoken words to written words are inadequate for transcribing video being used for research purposes. Programs designed for transcription, such Transana and InqScribe, can allow synchronization of video and transcript and are relatively inexpensive, but I used a method that requires no additional software. I transcribed by playing the video on the media player QuickTime, which has good video playing controls, while typing in a word processing program.

Because I had to watch each clip multiple times when I transcribed, transcribing my own clips provided me with the opportunity to become very familiar with the interactions and dialogue involved. Nevertheless, I have used other transcription options when I wanted a larger amount of data than I could transcribe on my own. I have used a professional transcription service for interviews, which are easier to transcribe because there are fewer speakers and the conversational turns rarely overlap. The cost of such a service is about $125- $150 per video hour and such transcriptions are usually completed faster than those done by a nonprofessional. Another option is to hire undergraduate students to transcribe. Undergraduate students can transcribe an hour of video in about six hours, and because their hourly rates are quite low, a student-generated transcript can cost half the price of a professional transcript. Physics undergraduates are often good transcribers because of their familiarity with the subject matter. I proofread the transcripts I had not generated myself, as people unfamiliar with physics or the particular situation misunderstand some terms and names.

### D. Processing video

Video shown at conferences often must be altered in some way. Two common requirements are captions and concealment of a face. Video captions can be created using Magpie 2, a free software program available for download from the National Center for Accessible Media. While Magpie is free and produces suitable captions, it is not particularly easy to use. Magpie requires the installation of the QuickTime player and the latest version of Java. An included help file lists detailed instructions on how to generate a file with captions synched with the video.

The least difficult way I found to obscure a portion of the screen was using Microsoft PowerPoint. I inserted the video into a slide and, using the Drawing toolbar, placed a box over a portion of the video in the slide. The box could be modified using the Format Shape menu option to be semi-translucent, if I desired.

Obscuring a portion of the screen could produce a distracting image. Another option was to modify the entire clip to have a cartoon appearance, which was done using QuickTime and Adobe's Photoshop program. I used the QuickTime command File/ Export and chose to export the sound track of the video to an .AIFF file. I then repeated the command, choosing the new option of exporting the video clip as a set of individual images. I modified one image using the 'Filter' option in Photoshop, and then saved the sequence of steps needed to do this as a small program (called an 'action' in Photoshop). This action could be automatically applied to all of the individual images. Finally, I recombined the modified images and the audio track in QuickTime. I did this by first selecting the command "Open Image Sequence" and choosing the set of modified images. I then opened the sound file in QuickTime, copied it, and pasted it into the image file using the Edit/Add menu command. I could have also used this method to blur a single face by using the filter option on only a portion of the screen.

### V. CONCLUSION

Collecting video to use in research can seem a daunting task. However, the use of consumer video products now available can diminish the technical hurdles. One camera and one microphone, costing only a few hundred dollars together, are sufficient equipment to begin taping, and video processing can be done with software commonly bundled with most computers. Using such consumer equipment can provide a more straightforward entry into videotaping in the classroom.

ACKNOWLEDGEMENTS

I am grateful for input from the Physics Education Research groups at the University of Colorado, Boulder and the University of Maryland, College Park, and I especially appreciate R. Scherr's thoughtful comments. This material is based upon work supported by the National Science Foundation under Grant No. NSF REC 0529482.